# Syntax and Semantics of Abstract Binding Trees

Jonathan Sterling and Darin Morrison

Abstract binding trees (abts) are a generalization of abstract syntax trees where operators may express variable binding structure as part of their arities. Originally formulated by Peter Aczel [1], unisorted abts have been deployed successfully as the uniform syntactic framework for several implementations of Constructive Type Theory, including Nuprl [3], MetaPRL [13] and JonPRL [21].

In *Practical Foundations for Programming Languages* [12], Robert Harper develops the multi-sorted version of abstract binding trees and proposes an extension to include families of operators indexed by *symbols*, which are, unlike variables, subject to only distinctness-preserving renaming and not substitution; furthermore, symbols do not appear in the syntax of abts, and are only introduced as parameters to operators. This extension to support symbols is essential for a correct treatment of programming languages with open sums (e.g. ML's `exn` type), as well as assignable references.

In parallel, M. Fiore and his collaborators have developed the categorical semantics for several variations of second-order algebraic theories [8, 10, 6, 7]; of these, the simply sorted variants are equivalent to Harper's abstract binding trees augmented with a notion of second-order variable (metavariable).

The contribution of this paper is the development of the syntax and semantics of multi-sorted nominal abts, an extension of second order universal algebra to support symbol-indexed families of operators. Additionally, we have developed the categorical semantics for abts formally in Constructive Type Theory using the Agda proof assistant [18]; we have also developed an implementation for abstract binding trees in Standard ML, which we intend to integrate into the JonPRL proof assistant [21].

## 1 Categorical Preliminaries

Fix a set $\mathcal{S}$ of *sorts*; we will say $\tau$ *sort* when $\tau \in \mathcal{S}$. In this section, we will develop a theory of sets varying over collections of $\mathcal{S}$-sorted symbols.

Let $\mathbb{I}$ be the category of finite cardinals and their injective maps; then the comma construction $\mathbb{I} \downarrow \mathcal{S}$, is the category of contexts of symbols whose objects are finite sets of symbols $\mathcal{U}$ and sort-assignments $\mathcal{U} \xrightarrow{\mathfrak{s}} \mathcal{S}$, and whose morphisms are sort-preserving renamings; we will write $\Upsilon$ for a symbol context $(\mathcal{U}, \mathfrak{s})$.

*Remark* 1.1. To be precise, $\mathbb{I} \downarrow \mathcal{S}$ is an abuse of notation for the comma category $|-|_{\mathbb{I}} \downarrow \Delta[\mathcal{S}]$, with $|-|_{\mathbb{I}}$ the forgetful functor from $\mathbb{I}$ to **Set**, and $\Delta[\mathcal{S}](*) \triangleq \mathcal{S}$ a constant functor from $\mathbb{1}$ to **Set**. Formally, an object of $\mathbb{I} \downarrow \mathcal{S}$ is, then, a triple $\langle U, *, \mathfrak{s} \rangle$ with $U$ an object in $\mathbb{I}$, $*$ the unique object in $\mathbb{1}$, and $\mathfrak{s}$ a function $|U|_{\mathbb{I}} \longrightarrow \mathcal{S}$; an arrow $\rho : \langle U, *, \mathfrak{s} \rangle \longrightarrow \langle V, *, t \rangle$ is a commuting triangle like the following:

$$\begin{array}{ccc} |U|_{\mathbb{I}} & \xrightarrow{|\rho|_{\mathbb{I}}} & |V|_{\mathbb{I}} \\ & {\scriptstyle \mathfrak{s}} \searrow \quad \swarrow {\scriptstyle t} & \\ & \mathcal{S} & \end{array}$$

Because the object $* : \mathbb{1}$ is unique, we will always omit it from the data of an object in $\mathbb{I} \downarrow \mathcal{S}$.

**Copresheaves on $\mathbb{I} \downarrow \mathcal{S}$** A presheaf on a category $\mathcal{C}$ is a contravariant functor $\mathcal{C}^{op} \longrightarrow$ **Set**; when $\mathcal{C} \equiv \mathcal{D}^{op}$, we call this a *copresheaf* (covariant presheaf) on $\mathcal{D}$. Therefore, a copresheaf on $\mathbb{I} \downarrow \mathcal{S}$ is a functor $\mathbb{I} \downarrow \mathcal{S} \longrightarrow$ **Set**; colloquially, such a copresheaf can be thought of as a set that *varies over* symbol contexts.

This leads to a very intuitive setting in which to consider sets that vary over collections of $\mathcal{S}$-sorted symbols. In particular, for any morphism $\Upsilon \overset{\rho}{\hookrightarrow} \Upsilon'$ in $\mathbb{I} \downarrow \mathcal{S}$, any copresheaf $X : \mathbf{Set}^{\mathbb{I} \downarrow \mathcal{S}}$ has a symbol-renaming action $X(\rho) : X(\Upsilon) \longrightarrow X(\Upsilon')$. When the copresheaf $X$ is clear from context, we will write $m \cdot \rho$ for $X(\rho)(m)$.

Now, when $m \in X(\Upsilon)$ for a copresheaf $X : \mathbf{Set}^{\mathbb{I} \downarrow \mathcal{S}}$, we can define the notion that $m$ depends on *at most* a subcontext $\Upsilon'$ of $\Upsilon$; this is called *support*.

**Definition 1.2** (Support)**.** When $\Upsilon \overset{\rho}{\hookrightarrow} \Upsilon'$, we say that $\Upsilon$ supports $m \in \Upsilon'$ (written $\Upsilon \blacktriangleright_\rho m$) when, for all $\Upsilon' \overset{\rho_1, \rho_2}{\hookrightarrow} \Upsilon''$, if $\rho_1 \circ \rho = \rho_2 \circ \rho$ then $m \cdot \rho_1 = m \cdot \rho_2$.

We can also always calculate the *least support* of a term $m \in X(\Upsilon')$, written $\mathbf{supp}(m)$:

$$\mathbf{supp}(m) \triangleq \bigcap_{\Upsilon \overset{\rho}{\hookrightarrow} \Upsilon'} \{ \Upsilon \mid \Upsilon \blacktriangleright_\rho m \}$$

Intuitively, $\mathbf{supp}(m)$ is the exact symbol context that $m$ depends on.

## 1.1 Sheaves on the atomic site

When $\Upsilon \overset{\rho}{\hookrightarrow} \Upsilon'$ such that $\Upsilon \blacktriangleright_\rho m$ for some $m \in X(\Upsilon')$, we would expect that we can work backward to a unique $m' : \Upsilon$ such that $m' \cdot \rho = m$; whilst it is not in general the case for presheaves $X$, this is precisely the sheaf condition for copresheaves on $\mathbb{I} \downarrow \mathcal{S}$ under the atomic topology [16, p. 126], which we will define in Definition 1.6.

The sheaf condition ensures that the notion of support is well-behaved in the following sense: if some term lies in two different fibers, then it may be *strengthened* uniquely into the fiber of its support.

**Definition 1.3** (Sieves and matching families)**.** A *sieve* $S$ on an object $C$ is a subfunctor of the Yoneda embedding $\mathbf{y}(C) \triangleq \mathcal{C}[-, C]$, i.e. $S \sqsubseteq \mathbf{y}(C)$. For a presheaf $X$ on the category $\mathcal{C}$, a *matching family* for a sieve $S$ is a natural transformation $\phi : S \longrightarrow X$. An *amalgamation* for $\phi$ is an object $m \in X(C)$ such that $m \cdot f = \phi_D(f)$ for each $f \in S(D)$, for all objects $D$.

**Definition 1.4** (Sites and coverages)**.** A *site* is a pair $\langle \mathcal{C}, J \rangle$, with $\mathcal{C}$ a category of "worlds" and $J$ a Grothendieck topology (or "coverage"). A Grothendieck topology $J$ (or *coverage*) is a family of sets of sieves, indexed by $\mathcal{C}$-objects; that is, for each object $C : \mathcal{C}$, $J(C)$ is the set of $C$-sieves which will be said to *cover* $C$. A topology $J$ must satisfy several axioms (see [16] for details).

**Definition 1.5** (Separation and sheafhood)**.** A presheaf on a site $\langle \mathcal{C}, J \rangle$ is a presheaf on $\mathcal{C}$; we say that such a presheaf $X$ is *separated* in case for every sieve $S \in J(C)$ and matching family $\phi : S \longrightarrow X$, there is at most one amalgamation $m \in X(C)$ for $\phi$. Additionally, $X$ is a *sheaf* when there is *exactly one* such amalgamation.

**Definition 1.6** (The atomic topology)**.** The atomic coverage $J_{\mathtt{atm}}$ is the one in which all non-empty sieves cover; in other words:

$$J_{\mathtt{atm}}(C) \triangleq \{ S \sqsubseteq \mathbf{y}(C) \mid \textstyle\bigcup_D S(D) \text{ non-empty} \}$$

In the future, we will write $\mathbb{I}[\mathcal{S}]$ for the Grothendieck site $\left\langle (\mathbb{I} \downarrow \mathcal{S})^{\mathrm{op}}, J_{\mathtt{atm}} \right\rangle$. Sheafhood on the atomic site $\mathbb{I}[\mathcal{S}]$ amounts to the following condition, using Definition 1.2:

> A copresheaf $X : \mathbf{Set}^{\mathbb{I}\downarrow\mathcal{S}}$ is a sheaf on $\mathbb{I}[\mathcal{S}]$ just when, for all $n \in X(\Upsilon')$, if we have $\Upsilon \xhookrightarrow{\rho} \Upsilon'$ and $\Upsilon \blacktriangleright_\rho n$, then there is a unique $m \in X(\Upsilon)$ such that $m \cdot \rho = n$.

Therefore, a sheaf on $\mathbb{I}[\mathcal{S}]$ is a set that varies over $\mathcal{S}$-sorted symbol contexts, which has a well-behaved notion of support. Another way to put it is, if $m \in X(\Upsilon)$ only "uses" some of the symbols available at world $\Upsilon$, there's a canonical way to restrict $m$ to a term in the world which contains *only* the symbols that it uses.

**Constructions on sheaves** Every category of sheaves on a site gives rise to a topos, which equips us with a number of standard constructions, including (among other things) the disjoint union of sheaves $X \oplus Y$, the product of sheaves $X \otimes Y$, and the terminal sheaf $\mathbb{1}$.

## 2 Operators and signatures

A valence $\{\vec{\sigma}\}[\vec{\tau}].\tau$ specifies an expression of sort $\tau$ which binds symbols in $\vec{\sigma}$ and variables in $\vec{\tau}$.

$$\frac{\tau\ \mathit{sort} \quad \sigma_i\ \mathit{sort}\ \ (i \leqslant m) \quad \tau_i\ \mathit{sort}\ \ (i \leqslant n)}{\{\sigma_0, \ldots, \sigma_m\}[\tau_0, \ldots, \tau_n].\tau\ \mathit{valence}}$$

An arity $(\vec{v})\,\tau$ specifies an operator of sort $\tau$ with arguments of valences $\vec{v}$. We will call the set of valences $\mathcal{V}_{\mathcal{S}}$, and the set of arities $\mathcal{A}_{\mathcal{S}}$.

$$\frac{\tau\ \textit{sort} \quad v_i\ \textit{valence}\ \ (i \leqslant n)}{(v_0, \ldots, v_n)\,\tau\ \textit{arity}}$$

Fix a family of sheaves of operators $\mathcal{O}_{\mathfrak{a}} \in \mathbf{Sh}\,(\mathbb{I}[\mathcal{S}])$, indexed by arities $\mathfrak{a} \in \mathcal{A}_{\mathcal{S}}$; for each $\mathcal{O}_{\mathfrak{a}}$, arrows in $\mathbb{I} \downarrow \mathcal{S}$ will lift to renamings in operators' symbolic parameters. Together, $\mathcal{S}$ and $\mathcal{O}$ are said to form a *signature* $\Sigma \triangleq \langle\, \mathcal{S}, \mathcal{O}\, \rangle$.

We will write $\Upsilon \Vdash \vartheta : \mathfrak{a}$ in case $\vartheta \in \mathcal{O}_{\mathfrak{a}}(\Upsilon)$; note that this judgment enjoys the structural properties of weakening and exchange via functoriality of $\mathcal{O}$, and strengthening via the sheaf condition. An operator signature is defined by specifying, for each arity $\mathfrak{a}$, the sheaf $\mathcal{O}_{\mathfrak{a}}$, whose "elements" are the operators of arity $\mathfrak{a}$.

**Examples** For instance, consider a $\lambda$-calculus with a single sort, exp; we give its signature $\Sigma_\lambda$ by asserting the following about its operators:

$$\Upsilon \Vdash \texttt{lam} : ([\texttt{exp}].\,\texttt{exp})\,\texttt{exp}$$
$$\Upsilon \Vdash \texttt{fix} : ([\texttt{exp}].\,\texttt{exp})\,\texttt{exp}$$
$$\Upsilon \Vdash \texttt{ap} : (.\,\texttt{exp}, .\,\texttt{exp})\,\texttt{exp}$$

These rules correspond to the following definition of $\mathcal{O}$:

$$\mathcal{O}_{([\texttt{exp}].\,\texttt{exp})\,\texttt{exp}} \cong \mathbb{1} \oplus \mathbb{1}$$
$$\mathcal{O}_{(.\,\texttt{exp}, .\,\texttt{exp})\,\texttt{exp}} \cong \mathbb{1}$$

So far, we have made no use of symbols and parameters; however, consider the extension of the calculus with assignables (references):

$$\Upsilon \Vdash \texttt{decl} : (.\,\texttt{exp}, \{\texttt{exp}\}.\,\texttt{exp})\,\texttt{exp}$$
$$\Upsilon, u : \texttt{exp} \Vdash \texttt{get}[u] : ()\,\texttt{exp}$$
$$\Upsilon, u : \texttt{exp} \Vdash \texttt{set}[u] : (.\,\texttt{exp})\,\texttt{exp}$$

Let $\mathbf{S}_\tau$ be the sheaf of symbols of sort $\tau$, a subobject of the Yoneda embedding of the empty symbol context, $\mathbf{y}(\cdot)$:

$$\mathbf{S}_\tau(\Upsilon) \triangleq \{\, u \in |\Upsilon| \mid \Upsilon(u) \equiv \tau \,\}$$

Then the above rules correspond to the following family of sheaves:

$$\mathcal{O}_{(.\,\texttt{exp}, \{\texttt{exp}\}.\,\texttt{exp})\,\texttt{exp}} \cong \mathbb{1}$$
$$\mathcal{O}_{()\,\texttt{exp}} \cong \mathbf{S}_{\texttt{exp}}$$
$$\mathcal{O}_{(.\,\texttt{exp})\,\texttt{exp}} \cong \mathbf{S}_{\texttt{exp}}$$

Declaring a new assignable consists in providing an initial value, and an expression binding a symbol (which shall represent the assignable in scope). Note that the functoriality of $\mathcal{O}$ guarantees for any renaming $\rho : \Upsilon \hookrightarrow \Upsilon'$, a family of lifted maps $\mathcal{O}_{\mathfrak{a}}(\rho) : \mathcal{O}_{\mathfrak{a}}(\Upsilon) \hookrightarrow \mathcal{O}_{\mathfrak{a}}(\Upsilon')$ natural in $\mathfrak{a}$, such that $\Upsilon' \Vdash \mathcal{O}(\rho)(\vartheta) : \mathfrak{a}$ when $\Upsilon \Vdash \vartheta : \mathfrak{a}$. In particular, the renaming $\Upsilon, u \mapsto \Upsilon, v$ shall take $\mathtt{get}[u]$ to $\mathtt{get}[v]$. In the future, we will write $\vartheta \cdot \rho$ for $\mathcal{O}_{\mathfrak{a}}(\rho)(\vartheta)$.

## 3 Contexts

In general, we will have three kinds of context: symbolic (parameter) contexts, variable contexts, and metavariable contexts. The symbol contexts have already been defined via the comma construction $\mathbb{I} \downarrow \mathcal{S}$, but they also admit a syntactic characterization,

$$\frac{}{\cdot \ \mathit{sctx}} \qquad \frac{\Upsilon \ \mathit{sctx} \quad \tau \ \mathit{sort} \quad u \notin |\Upsilon|}{\Upsilon, u : \tau \ \mathit{sctx}}$$

Because, modulo notation, we have $\Upsilon \in \mathbb{I} \downarrow \mathcal{S}$ just when $\Upsilon$ *sctx*, we will use the syntactic view when it is convenient.

Contexts of variables are similar to contexts of symbols, except that they admit *any* renamings, not just the injective ones. As such, when $\mathbb{F}$ is the category of finite cardinals and all functions between them, the comma construction $\mathbb{F} \downarrow \mathcal{S}$ is the category of variable contexts. As above, we can give them an equivalent syntactic treatment:

$$\frac{}{\cdot \ \mathit{vctx}} \qquad \frac{\Gamma \ \mathit{vctx} \quad \tau \ \mathit{sort} \quad x \notin |\Gamma|}{\Gamma, x : \tau \ \mathit{vctx}}$$

A metavariable context consists of bindings of $\mathcal{S}$-*valences* to metavariables; let $\mathcal{V}_{\mathcal{S}} \triangleq \{ v \mid v \ \mathit{valence} \}$ be the set of valences. Then, the category of metavariable contexts is the comma construction $\mathbb{F} \downarrow \mathcal{V}_{\mathcal{S}}$, which likewise admits an equivalent inductive definition:

$$\frac{}{\cdot \ \mathit{mctx}} \qquad \frac{\Theta \ \mathit{mctx} \quad v \ \mathit{valence} \quad \mathfrak{m} \notin |\Theta|}{\Theta, \mathfrak{m} : v \ \mathit{mctx}}$$

## 4 Nominal Abstract Binding Trees

Let the judgment $\Theta \rhd \Upsilon \parallel \Gamma \vdash M : \tau$ presuppose [1] $\Theta$ *mctx*, $\Upsilon$ *sctx*, $\Gamma$ *vctx* and $\tau$ *sort*, meaning that $M$ is an abstract binding tree of sort $s$, with metavariables in $\Theta$, parameters in $\Upsilon$, and variables in $\Gamma$. Let the judgment $\Theta \triangleright \Upsilon \parallel \Gamma \vdash E : \nu$ presuppose $\nu$ *valence*. Then, the syntax of abstract binding trees is inductively defined in four rules:

$$\frac{\Gamma \ni x : \tau}{\Theta \triangleright \Upsilon \parallel \Gamma \vdash x : \tau} \vdash_{var}$$

$$\frac{\begin{array}{l}\Theta \ni \mathfrak{m} : \{\sigma_0, \dots, \sigma_m\}[\tau_0, \dots, \tau_n].\tau \\ \Upsilon \ni u_i : \sigma_i \quad (i \leqslant m) \\ \Theta \triangleright \Upsilon \parallel \Gamma \vdash M_i : \tau_i \quad (i \leqslant n)\end{array}}{\Theta \triangleright \Upsilon \parallel \Gamma \vdash \mathfrak{m}\{u_0, \dots, u_m\}(M_0, \dots, M_n) : \tau} \vdash_{mvar}$$

$$\frac{\begin{array}{l}\Upsilon \Vdash \vartheta : (\nu_1, \dots, \nu_n)\tau \\ \Theta \triangleright \Upsilon \parallel \Gamma \vdash E_i : \nu_i \quad (i \leqslant n)\end{array}}{\Theta \triangleright \Upsilon \parallel \Gamma \vdash \vartheta(E_0, \dots, E_n) : \tau} \vdash_{app}$$

$$\frac{\Theta \triangleright \Upsilon, \vec{u} : \vec{\sigma} \parallel \Gamma, \vec{x} : \vec{\tau} \vdash M : \tau}{\Theta \triangleright \Upsilon \parallel \Gamma \vdash \lambda\!\!\lambda\{\vec{u}\}[\vec{x}].M : \{\vec{\sigma}\}[\vec{\tau}].\tau} \vdash_{abs}$$

[1] In the *judgmental method*, as pioneered by Per Martin-Löf, a form of judgment is propounded first by specifying its range of significance (i.e. the circumstances under which it shall have a meaning), and then, a meaning explanation (definition) for the judgment is given. The range of significance of a judgment is called its *presupposition*, and the meaning of the judgment may proceed by induction on the evidence that the presupposition obtains. See [20] for a more detailed explanation of presuppositions and the judgmental method.

Abts are identified up to $\alpha$-equivalence.

## 4.1 Calculating free variables

We can easily calculate the variables free in an expression by recursion on its structure:

$$\frac{}{\mathbf{FV}(x) \rightsquigarrow \{x\}} \mathbf{FV}_{var}$$

$$\frac{\mathbf{FV}(M_i) \rightsquigarrow \vec{x}_i \quad (i \leqslant n)}{\mathbf{FV}(\mathfrak{m}\{\vec{u}\}(M_0, \dots, M_n)) \rightsquigarrow \bigcup_{i \leqslant n} \vec{x}_i} \mathbf{FV}_{mvar}$$

$$\frac{\mathbf{FV}(E_i) \rightsquigarrow \vec{x}_i \quad (i \leqslant n)}{\mathbf{FV}(\vartheta(E_0, \dots, E_n)) \rightsquigarrow \bigcup_{i \leqslant n} \vec{x}_i} \mathbf{FV}_{app}$$

$$\frac{\mathbf{FV}(M) \rightsquigarrow \vec{x}}{\mathbf{FV}(\lambda\!\!\lambda\{\vec{u}\}[\vec{y}].M) \rightsquigarrow \vec{x} \setminus \vec{y}} \mathbf{FV}_{abs}$$

Because this is a total relation, henceforth we will write $\mathbf{FV}(M)$ for $\vec{x}$ when $\mathbf{FV}(M) \rightsquigarrow \vec{x}$.

## 4.2 Calculating free symbols

Whereas the calculation of free variables pivoted on the $\mathbf{FV}_{var}$ rule, the calculation of free symbols will pivot on the $\mathbf{FS}_{app}$ and $\mathbf{FS}_{mvar}$ rules, because the only way a symbol can be introduced is as a parameter to an operator or as a parameter to a metavariable.

$$\frac{}{\mathbf{FS}\,(x) \rightsquigarrow \{\,\}}\ \mathbf{FS}_{\mathit{var}}$$

$$\frac{\mathbf{FS}\,(M_i) \rightsquigarrow \vec{u}_i \quad (i \leqslant n)}{\mathbf{FS}\,(\mathfrak{m}\{\vec{u}\}(M_0, \ldots, M_n)) \rightsquigarrow \vec{u} \cup \bigcup_{i \leqslant n} \vec{u}_i}\ \mathbf{FS}_{\mathit{mvar}}$$

$$\frac{\mathbf{FS}\,(E_i) \rightsquigarrow \vec{u}_i \quad (i \leqslant n)}{\mathbf{FS}\,(\vartheta(E_0, \ldots, E_n)) \rightsquigarrow |\mathbf{supp}(\vartheta)| \cup \bigcup_{i \leqslant n} \vec{u}_i}\ \mathbf{FS}_{\mathit{app}}$$

$$\frac{\mathbf{FS}\,(M) \rightsquigarrow \vec{u}}{\mathbf{FS}\,(\lambda\!\!\lambda\{\vec{v}\}[\vec{x}].\, M) \rightsquigarrow \vec{u} \setminus \vec{v}}\ \mathbf{FS}_{\mathit{abs}}$$

Because this is a total relation, henceforth we will write $\mathbf{FS}\,(M)$ for $\vec{u}$ when $\mathbf{FS}\,(M) \rightsquigarrow \vec{u}$.

## 4.3 Renaming of symbols

The only place that symbols appear in our calculus is as parameters to operators (unlike variables, symbols are not terms). Therefore, the functorial action of the operator sheaf can be lifted into terms by recursion on their structure, via a pair of judgments $M \cdot \rho \rightsquigarrow N$ and $E \cdot \rho \rightsquigarrow F$, presupposing $\rho : \Upsilon \hookrightarrow \Upsilon'$, and $\Theta \triangleright \Upsilon \parallel \Gamma \vdash M : \tau$ and $\Theta \triangleright \Upsilon \parallel \Gamma \vdash E : \nu$ respectively:

$$\frac{}{x \cdot \rho \rightsquigarrow x}\ \bullet_{\mathit{var}}$$

$$\frac{\rho(\vec{u}) \equiv \vec{v} \quad M_i \cdot \rho \rightsquigarrow N_i \quad (i \leqslant n)}{\mathfrak{m}\{\vec{u}\}(M_0, \ldots, M_n) \cdot \rho \rightsquigarrow \mathfrak{m}\{\vec{v}\}(N_0, \ldots, N_n)}\ \bullet_{\mathit{mvar}}$$

$$\frac{E_i \cdot \rho \rightsquigarrow F_i \quad (i \leqslant n)}{\vartheta(E_0, \ldots, E_n) \cdot \rho \rightsquigarrow \vartheta \cdot \rho(F_0, \ldots, F_n)}\ \bullet_{\mathit{app}}$$

$$\frac{M \cdot \rho \setminus \vec{u} \rightsquigarrow N}{\lambda\!\!\lambda\{\vec{u}\}[\vec{x}].\, M \cdot \rho \rightsquigarrow \lambda\!\!\lambda\{\vec{u}\}[\vec{x}].\, N}\ \bullet_{\mathit{abs}}$$

Above, the notation $\rho \setminus \vec{u}$ means the omission of the variables $\vec{u}$ from the renaming $\rho$. Because terms are identified up to $\alpha$-equivalence, the renaming judgment is functional in its input, and so we are justified in writing $M \cdot \rho$ for $N$ when $M \cdot \rho \rightsquigarrow N$.

## 4.4 Substitution of variables

Variable substitution in abts is defined inductively by a pair of judgments, $[N \,/\, x]\, M \rightsquigarrow M'$ and $[N \,/\, x]\, E \rightsquigarrow F$:

$$\frac{x = y}{[N \,/\, x]\, y \rightsquigarrow N}\ /var_1 \qquad \frac{x \mathbin{\#} y}{[N \,/\, x]\, y \rightsquigarrow y}\ /var_2$$

$$\frac{[N \,/\, x]\, M_i \rightsquigarrow M'_i \quad (i \leqslant n)}{[N \,/\, x]\, \mathfrak{m}\{\vec{u}\}(M_0, \ldots, M_n) \rightsquigarrow \mathfrak{m}\{\vec{u}\}(M'_0, \ldots, M'_n)}\ /mvar$$

$$\frac{[N \,/\, x]\, E_i \rightsquigarrow F_i \quad (i \leqslant n)}{[N \,/\, x]\, \vartheta(E_0, \ldots, E_n) \rightsquigarrow \vartheta(F_0, \ldots, F_n)}\ /app$$

$$\frac{x \notin \vec{y} \quad \vec{u} \mathbin{\#} \mathbf{FS}(N) \quad \vec{y} \mathbin{\#} \mathbf{FV}(N) \quad [N \,/\, x]\, M \rightsquigarrow M'}{[N \,/\, x]\, \lambda\{\vec{u}\}[\vec{y}].\, M \rightsquigarrow \lambda\{\vec{u}\}[\vec{y}].\, M'}\ /abs_1$$

$$\frac{x \in \vec{y} \quad \vec{u} \mathbin{\#} \mathbf{FS}(N) \quad \vec{y} \mathbin{\#} \mathbf{FV}(N)}{[N \,/\, x]\, \lambda\{\vec{u}\}[\vec{y}].\, M \rightsquigarrow \lambda\{\vec{u}\}[\vec{y}].\, M}\ /abs_2$$

In the remainder, we will write $[N \,/\, x]\, M$ for $M'$ when $[N \,/\, x]\, M \rightsquigarrow M'$, and $\left[\vec{N} \,/\, \vec{x}\right] M$ for the simultaneous substitution of $\vec{N}$ for $\vec{x}$ in $M$.

### 4.5 Substitution of metavariables

Metavariables are substituted by abstractions; since a metavariable may only appear in an application expression $\mathfrak{m}\{\cdots\}(\cdots)$, we will instantiate the abstraction at the supplied parameters and arguments. Substitution for metavariables is defined inductively by the judgments $[E \,/\, \mathfrak{m}]\, M \rightsquigarrow N$ and $[E \,/\, \mathfrak{m}]\, F \rightsquigarrow F'$:

$$\frac{}{[E \,/\, \mathfrak{m}]\, x \rightsquigarrow x}\ /{}^{\texttt{meta}}_{var}$$

$$\frac{\mathfrak{m} \mathbin{\#} \mathfrak{n} \quad [E \,/\, \mathfrak{n}]\, M_i \rightsquigarrow N_i \quad (i \leqslant n)}{[E \,/\, \mathfrak{m}]\, \mathfrak{n}\{\vec{u}\}(M_0, \ldots, M_n) \rightsquigarrow \mathfrak{n}\{\vec{u}\}(N_0, \ldots, N_n)}\ /{}^{\texttt{meta}}_{mvar_1}$$

$$\frac{\mathfrak{m} = \mathfrak{n} \quad \left[\vec{M} \,/\, \vec{x}\right] N \rightsquigarrow N' \quad N' \cdot \vec{u} \mapsto \vec{v} \rightsquigarrow N''}{[\lambda\{\vec{u}\}[\vec{x}].\, N \,/\, \mathfrak{m}]\, \mathfrak{n}\{\vec{v}\}(\vec{M}) \rightsquigarrow N''}\ /{}^{\texttt{meta}}_{mvar_2}$$

$$\frac{[E \,/\, \mathfrak{m}]\, F_i \rightsquigarrow F'_i \quad (i \leqslant n)}{[E \,/\, \mathfrak{m}]\, \vartheta(F_0, \ldots, F_n) \rightsquigarrow \vartheta(F'_0, \ldots, F'_n)}\ /{}^{\texttt{meta}}_{app}$$

$$\frac{\vec{u} \mathbin{\#} \mathbf{FS}(E) \quad \vec{x} \mathbin{\#} \mathbf{FV}(E) \quad [E \,/\, \mathfrak{m}]\, M \rightsquigarrow N}{[E \,/\, \mathfrak{m}]\, \lambda\{\vec{u}\}[\vec{x}].\, M \rightsquigarrow \lambda\{\vec{u}\}[\vec{x}].\, N}\ /{}^{\texttt{meta}}_{abs}$$

The rules for metavariable substitution are reminiscent of the "hereditary substitutions" first devised for the Concurrent Logical Framework by Watkins in [22]. As usual, we will write $[E \,/\, \mathfrak{m}]\, M$ for $N$ when $[E \,/\, \mathfrak{m}]\, M \rightsquigarrow N$.

# 5 Model Theory

Let $\mathbf{H} \triangleq (\mathbb{I} \downarrow \mathcal{S} \times \mathbb{F} \downarrow \mathcal{S})^{\mathsf{op}}$; then we fix the functor category $\widehat{\mathbf{H}}^{\mathcal{S}}$ as our semantic universe, using the notation of the French school, $\widehat{\mathcal{C}} \triangleq \mathbf{Set}^{\mathcal{C}^{\mathsf{op}}}$. Let $\mathbf{V}_\tau(\Upsilon \parallel \Gamma) \triangleq \{x \in |\Gamma| \mid \Gamma(x) \equiv \tau\}$ be called the presheaf of variables; additionally, we have a presheaf of symbols $\mathbf{S}_\tau(\Upsilon \parallel \Gamma) \triangleq \{u \in |\Upsilon| \mid \Upsilon(u) \equiv \tau\}$. Lastly, we have the presheaf of operators with arity $\alpha$, $\mathcal{O}_\alpha(\Upsilon \parallel \Gamma) \triangleq \mathcal{O}_\alpha(\Upsilon)$

## 5.1 Substitution monoidal structures

For an object $P : \widehat{\mathbf{H}}^{\mathcal{S}}$, we will use the notation $P^{[\Gamma]}$ to mean $\prod_{x \in |\Gamma|} P_{\Gamma(x)}$; likewise, $\mathbf{S}^{\{\Upsilon\}}$ shall mean $\prod_{u \in |\Upsilon|} S_{\Upsilon(u)}$. For a presheaf $A : \widehat{\mathbf{H}}$ and a sort-indexed family of presheaves $P : \widehat{\mathbf{H}}^{\mathcal{S}}$, we have an operation $A \bullet P$, defined as a coend in the following way:

$$(A \bullet P)(\Upsilon \parallel \Gamma) \triangleq \int^{(\Upsilon' \parallel \Delta) \in \mathbf{H}} A(\Upsilon' \parallel \Delta) \times \mathbf{S}^{\{\Upsilon'\}}(\Upsilon \parallel \Gamma) \times P^{[\Delta]}(\Upsilon \parallel \Gamma)$$

Intuitively, the coend construction $A \bullet P$ provides the data of a suspended substitution of an $A$-term's variables by $P$-terms. Using this, we can define a tensor $P \odot Q$ for $P, Q : \widehat{\mathbf{H}}^{\mathcal{S}}$ as follows:

$$(P \odot Q)_\tau \triangleq P_\tau \bullet Q \qquad (\tau \in \mathcal{S})$$

Then, $V$ is the unit to this tensor. We will say that an object $P : \widehat{\mathbf{H}}^{\mathcal{S}}$ is a $\Sigma$-monoid in case it is equipped with the following natural transformations where $\nu$ embeds variables into $P$ and $\varsigma$ equips $P$ with an operation for simultaneous substitutions of variables. Furthermore, $\nu$ and $\varsigma$ induce maps $\nu_\Gamma$ and $\varsigma^\tau_{\Upsilon \parallel \Gamma}$:

$$V \xrightarrow{\ \nu\ } P \xleftarrow{\ \varsigma\ } P \odot P$$

$$V^{[\Gamma]} \xrightarrow{\ \nu_\Gamma\ } P^{[\Gamma]} \qquad P_\tau^{\mathbf{y}(\Upsilon \parallel \Gamma)} \times \mathbf{S}^{\{\Upsilon\}} \times P^{[\Gamma]} \xrightarrow{\ \varsigma^\tau_{\Upsilon \parallel \Gamma}\ } P_\tau$$

## 5.2 The signature endofunctor and its algebras

For each signature $\Sigma \equiv \langle \mathcal{S}, \mathcal{O} \rangle$, we have an endofunctor $\mathcal{F}_\Sigma : \widehat{\mathbf{H}}^{\mathcal{S}} \longrightarrow \widehat{\mathbf{H}}^{\mathcal{S}}$, which is defined as follows:

$$\mathcal{F}_\Sigma(X)_\tau \triangleq \coprod_{\vartheta \in \mathcal{O}_{(\vec{v})\tau}} \prod_{\{\vec{\sigma}\}[\vec{\tau}].\ \tau_i \in \vec{v}} X_{\tau_i}^{\mathbf{y}(\vec{\sigma} \parallel \vec{\tau})}$$

Then, a $\Sigma$-model is a $\Sigma$-monoid $P$ which is equipped with an algebra $\alpha : \mathcal{F}_\Sigma(P) \longrightarrow P$, which shall interpret applications of each operator.

### 5.3 Interpretation of terms

The metavariable, symbol and variable contexts are interpreted for a model $P$ as an environment presheaf in the following way:

$$[\![\Theta \triangleright \Upsilon \parallel \Gamma]\!]_P \triangleq \left( \prod_{(\mathfrak{m}:\{\vec{\sigma}\}[\vec{\tau}].\,\tau)\in\Theta} {P_\tau}^{\mathbf{y}(\vec{\sigma}\parallel\vec{\tau})} \right) \times \mathbf{S}^{\{\Upsilon\}} \times \mathbf{V}^{[\Gamma]}$$

Then, the interpretation of a term in a model $P$ is a map from its environment to $P$:

$$[\![\Theta \triangleright \Upsilon \parallel \Gamma \vdash M : \tau]\!]_P : [\![\Theta \triangleright \Upsilon \parallel \Gamma]\!]_P \longrightarrow P_\tau$$

Variables are interpreted by the map $[\![\Theta \triangleright \Upsilon \parallel \Gamma \vdash x : \tau]\!]_P$ which projects them from the environment and embeds them into the model. Metavariables are resolved by the map $\left[\!\!\left[\Theta \triangleright \Upsilon \parallel \Gamma \vdash \mathfrak{m}\{\vec{u}\}(\vec{M}) : \tau\right]\!\!\right]_P$ (where $\Theta \ni \mathfrak{m} : \{\vec{\sigma}\}[\vec{\tau}].\,\tau$) which projects their interpretation from the environment and instantiates it via substitution:

$$P_\tau \xleftarrow{[\![\Theta\triangleright\Upsilon\parallel\Gamma\vdash\mathfrak{m}\{\vec{u}\}(\vec{M}):\tau]\!]_P} [\![\Theta \triangleright \Upsilon \parallel \Gamma]\!]_P \xrightarrow{[\![\Theta\triangleright\Upsilon\parallel\Gamma\vdash x:\tau]\!]_P} P_\tau$$

$$[\![\Theta \triangleright \Upsilon \parallel \Gamma]\!]_P \xrightarrow{\langle \pi_\mathfrak{m}\,\pi_1, \phi\,\pi_2, \psi\rangle} P_\tau^{\mathbf{y}(\vec{\sigma}\parallel\vec{\tau})} \times \mathbf{S}^{\{\vec{\sigma}\}} \times P^{[\vec{\tau}]} \xrightarrow{\varsigma^\tau_{\vec{\sigma}\parallel\vec{\tau}}} P_\tau$$

$$[\![\Theta \triangleright \Upsilon \parallel \Gamma]\!]_P \xrightarrow{\nu_\Gamma\,\pi_3} P^{[\Gamma]} \xrightarrow{\pi_x} P_\tau$$

$$P^{[\vec{\tau}]} \xleftarrow{\psi \triangleq \left\langle [\![\Theta\triangleright\Upsilon\parallel\Gamma\vdash M:\tau]\!]_P \right\rangle_{(M,\tau)\in(\vec{M},\vec{\tau})}} [\![\Theta \triangleright \Upsilon \parallel \Gamma]\!]_P \qquad \mathbf{S}^{\{\Upsilon\}} \xrightarrow{\phi \triangleq \langle \pi_u \rangle_{u\in\vec{u}}} \mathbf{S}^{\{\vec{\sigma}\}}$$

Interpretation of operator applications is the most complicated. Recall that, unlike in standard treatments of universal algebra, our operators are indexed by symbol collections; therefore, operators must pass through suitable renamings in order to be used in the interpretation. Let us begin by constructing for each operator $\Upsilon \Vdash \vartheta : \mathfrak{a}$ the morphism $[\![\vartheta]\!]_P$ which shall rename the parameters of the operator using the environment:

$$[\![\vartheta]\!]_P : [\![\Theta \triangleright \Upsilon \parallel \Gamma]\!]_P \xrightarrow{\pi_2} \mathbf{S}^{\{\Upsilon\}} \xrightarrow{\mathcal{O}_\mathfrak{a}(\pi_{(-)})(\vartheta)} \mathcal{O}_\mathfrak{a}$$

We will proceed using the $\mathcal{F}_\Sigma$-algebra $\alpha$, as follows, by composing it with a morphism $\beta$ from the environment into the signature endofunctor, which interprets the syntax of operator applications:

$$[\![\Theta \triangleright \Upsilon \parallel \Gamma]\!]_P \xrightarrow{\beta} \mathcal{F}_\Sigma(P)_\tau \xrightarrow{\alpha_\tau} P_\tau$$

The construction of $\beta$ proceeds by renaming the parameters of the operator $\vartheta$ and constructing the (bound) exponentiated arguments $\gamma$ of the operator.

$$[\![\Theta \triangleright \Upsilon \parallel \Gamma]\!]_P \xrightarrow{\beta \triangleq \langle [\![\vartheta]\!]_P, \lambda\gamma \rangle} \coprod_{\vartheta \in \mathcal{O}_{(\vec{v})\tau}} \prod_{\{\vec{\sigma}_i\}[\vec{\tau}_i].\tau_i \in \vec{v}} P_{\tau_i}^{\mathbf{y}(\vec{\sigma}_i \parallel \vec{\tau}_i)}$$

Arguments $\lambda\gamma_i$ are the exponential transposes (curried form) of the composites $\gamma_i$:

$$[\![\Theta \triangleright \Upsilon \parallel \Gamma]\!]_P \xrightarrow{\lambda\gamma_i} P_{\tau_i}^{\mathbf{y}(\vec{\sigma}_i \parallel \vec{\tau}_i)}$$

$$[\![\Theta \triangleright \Upsilon \parallel \Gamma]\!]_P \times \mathbf{y}(\vec{\sigma}_i \parallel \vec{\tau}_i) \xrightarrow{\langle \pi_{[1,1]}, \phi_i, \psi_i \rangle} [\![\Theta \triangleright \Upsilon, \vec{\sigma}_i \parallel \Gamma, \vec{\tau}_i]\!]_P \xrightarrow{[\![\Theta \triangleright \Upsilon, \vec{\sigma}_i \parallel \Gamma, \vec{\tau}_i \vdash M_i : \tau_i]\!]_P} P_{\tau_i}$$

$$[\![\Theta \triangleright \Upsilon \parallel \Gamma]\!]_P \times \mathbf{y}(\vec{\sigma}_i \parallel \vec{\tau}_i) \xrightarrow{\gamma_i} P_{\tau_i}$$

where $\phi_i, \psi_i$ are defined as follows:

$$[\![\Theta \triangleright \Upsilon \parallel \Gamma]\!]_P \times \mathbf{y}(\vec{\sigma}_i \parallel \vec{\tau}_i) \xrightarrow{\phi_i \triangleq \langle \pi_u \circ \pi_1 \rangle_{u \in |\Upsilon|}, \langle \pi_u \circ \pi_{[2,1]} \rangle_{u \in |\vec{\sigma}_i|}} \mathbf{S}^{\{\Upsilon, \vec{\sigma}_i\}}$$

$$[\![\Theta \triangleright \Upsilon \parallel \Gamma]\!]_P \times \mathbf{y}(\vec{\sigma}_i \parallel \vec{\tau}_i) \xrightarrow{\psi_i \triangleq \langle \pi_x \circ \pi_1 \rangle_{x \in |\Gamma|}, \langle \pi_x \circ \pi_{[2,1]} \rangle_{x \in |\vec{\tau}_i|}} \mathbf{V}^{[\Gamma, \vec{\tau}_i]}$$

This concludes the interpretation of well-sorted terms into any $\Sigma$-model.

## 6 Case Study: Wellformed Sequents

The representation of telescopes and sequents in a logical framework is notoriously difficult; whilst it is possible to use higher-order abstract syntax or abts to encode the binding-structure of telescopes and sequents, the encoding is sufficiently laborious and obscure that it is not used in practice. Crary has demonstrated a first-order encoding of contexts in the Edinburgh Logical Framework (ELF) in bijection with actual LF-contexts [4], which has been successfully used in large-scale mechanization efforts, including that of Standard ML [15] and the ELF itself [17].

We will approach the problem of encoding telescopes and sequents from the *refinements* perspective, where a conservative approximation of the grammar is first given using the abt logical framework, and then the correctness of a code is expressed separately in a judgment that refines the existing specification. Because we have not committed to using the built-in binding machinery to express the well-scopedness of telescopes and sequents, we are free to use *symbols* in order to model the variables in the context. This is in fact quite sensible if we are actually trying to faithfully represent the syntax of telescopes and sequents, rather than replace them with their counterparts on the meta-level.

The approach outlined above is actually quite similar to the standard practice in the Abella proof assistant [11]; in Abella, in addition to the canonical forms, each type is inhabited by an infinite collection of nominal constants or symbols, which can be used in the syntax of contexts. In order to specify where such symbols are allowed to occur, a wellformedness predicate (a "scheme") is separately defined. Our setting is similar in that we may also use symbols to encode the syntax of contexts, except that here, we have better control over their proliferation because, contrary to the state of affairs in Abella, symbols are *not* expressions; following Harper [12], symbols are meant only to serve as an open-ended indexing scheme for operators, not as actual expressions.

There are two reasons that one might want to model sequents in a (meta)-logical framework. Perhaps the most obvious reason is to prove things about a logic with sequent judgments; this is what motivated Crary's encoding of contexts in LF, for instance. However, another common use-case is the practical development of a proof assistant in which users can define new sequent rules; such user-defined rules may only be compiled into well-behaved tactics in the proof system in case they obey certain wellformedness properties including (among other things) correct scoping of names. MetaPRL is a salient example of such a proof assistant, whose implementation might have benefited from the more principled treatment of symbols and metavariables which we present here.

This insight leads the way to a simple abt signature for the theory of telescopes and sequents.

| | |
|---|---|
| $\mathtt{tele}$ *sort* | *telescopes* |
| $\mathtt{exp}$ *sort* | *expressions* |
| $\mathtt{prop}$ *sort* | *propositions* |
| $\mathtt{jdg}$ *sort* | *judgments* |

$$\Upsilon, u : \mathtt{exp} \Vdash \mathtt{hyp}[u] : ()\, \mathtt{exp}$$

$$\Upsilon \Vdash \mathtt{nil} : ()\, \mathtt{tele}$$
$$\Upsilon, u : \mathtt{exp} \Vdash \mathtt{snoc}[u] : (.\mathtt{tele}, .\mathtt{prop})\, \mathtt{tele}$$

$$\Upsilon \Vdash \mathtt{sequent} : (.\mathtt{tele}, .\mathtt{prop})\, \mathtt{jdg}$$

Suppose we have encoded a fragment of type theory as well:

$$\Upsilon \Vdash \mathtt{P} : ()\, \mathtt{prop}$$
$$\Upsilon \Vdash \mathtt{pred} : (.\mathtt{exp})\, \mathtt{prop}$$

Terms written using the abstract syntax will be difficult to read, so let us define some notation:

$$\diamond \triangleq \mathtt{nil}$$
$$H, u : P \triangleq \mathtt{snoc}[u](H, P)$$
$$H \gg A \triangleq \mathtt{sequent}(H, A)$$
$$\lq u \triangleq \mathtt{hyp}[u]$$

Now, we can write the following sequent:

$$\cdot \triangleright u : \mathtt{exp}, v : \mathtt{exp} \parallel \cdot \vdash \diamond, u : \mathtt{P}, v : \mathtt{pred}(\text{‘}u) \gg \mathtt{pred}(\text{‘}u) : \mathtt{jdg}$$

The above sequent has free symbols, but we can close over them by adding a form of parametric higher-order judgment (as in [12]) to our object language, indexed by a collection of sorts $\vec{\sigma}$:

$$\Upsilon \Vdash \nabla[\vec{\sigma}] : (\{\vec{\sigma}\}.\,\mathtt{jdg})\,\mathtt{jdg}$$

Then, we may write a closed sequent judgment as follows:

$$\cdot \triangleright \cdot \parallel \cdot \vdash \nabla[\mathtt{exp}, \mathtt{exp}](\lambda\{u, v\}[].\,\diamond, u : \mathtt{P}, v : \mathtt{pred}(\text{‘}u) \gg \mathtt{pred}(\text{‘}u)) : \mathtt{jdg}$$

## 6.1 Refinements for wellformedness

Having specified an approximation of the grammar of telescopes and sequents in the abt logical framework, we can proceed to define proper wellformedness via *inductive refinement* [12]. The basic idea is to introduce a new form of (meta)-judgment $\Upsilon \parallel \Gamma \vdash M \in_{\mathbf{wf}} \tau$ which expresses the extrinsic wellformedness properties we wish to verify, presupposing $\cdot \triangleright \Upsilon \parallel \Gamma \vdash M : \tau$. Additionally, we introduce an analogous judgment on abstractions, $\Upsilon \parallel \Gamma \vdash E \in_{\mathbf{wf}} \nu$ presupposing $\cdot \triangleright \Upsilon \parallel \Gamma \vdash E : \nu$, defined uniformly as follows:

$$\frac{\Upsilon, \vec{u} : \vec{\sigma} \parallel \Gamma, \vec{x} : \vec{\tau} \vdash M \in_{\mathbf{wf}} \tau}{\Upsilon \parallel \Gamma \vdash \lambda\{\vec{u}\}[\vec{x}].\,M \in_{\mathbf{wf}} \{\vec{\sigma}\}[\vec{\tau}].\,\tau}$$

Likewise, wellformedness for variables is defined uniformly:

$$\frac{}{\Upsilon \parallel \Gamma \vdash x \in_{\mathbf{wf}} \tau}$$

Note that the refinement for variables $x$ is not trivial, since it is only defined in case the presupposition $\cdot \triangleright \Upsilon \parallel \Gamma \vdash x : \tau$ is satisfied.

The remainder of the definition of refinement proceeds by induction on sorts and operators. For the sake of this example, we will just stipulate that anything of sort `exp` or `prop` is grammatical if its subterms are grammatical:

$$\frac{\Upsilon \Vdash \vartheta : (\nu_0, \ldots, \nu_n)\,\tau \qquad \Upsilon \parallel \Gamma \vdash E_i \in_{\mathbf{wf}} \nu_i \;\; (i \leqslant n)}{\Upsilon \parallel \Gamma \vdash \vartheta(E_0, \ldots, E_n) \in_{\mathbf{wf}} \tau} \text{ for } \tau \in \{\mathtt{exp}, \mathtt{prop}\}$$

The refinements for parametric judgment and sequents simply delegate to their subterms as well:

$$\frac{\Upsilon, \vec{u} : \vec{\sigma} \parallel \Gamma \vdash J \in_{\mathbf{wf}} \mathtt{jdg}}{\Upsilon \parallel \Gamma \vdash \nabla[\vec{\sigma}](\lambda\{\vec{u}\}[].\,J) \in_{\mathbf{wf}} \mathtt{jdg}}$$

$$\frac{\Upsilon \parallel \Gamma \vdash H \in_{\mathbf{wf}} \mathtt{tele} \qquad \Upsilon \parallel \Gamma \vdash A \in_{\mathbf{wf}} \mathtt{prop}}{\Upsilon \parallel \Gamma \vdash H \gg A \in_{\mathbf{wf}} \mathtt{jdg}}$$

The refinement for telescopes proceeds by induction:

$$\frac{}{\Upsilon \parallel \Gamma \vdash \diamond \in_{\mathbf{wf}} \texttt{tele}} \qquad \frac{\Upsilon \setminus \{u\} \parallel \Gamma \vdash H \in_{\mathbf{wf}} \texttt{tele} \quad \Upsilon \setminus \{u\} \parallel \Gamma \vdash A \in_{\mathbf{wf}} \texttt{prop}}{\Upsilon \parallel \Gamma \vdash H, u : A \in_{\mathbf{wf}} \texttt{tele}}$$

Based on these rules, it is easy to show that the example sequent from the previous section is wellformed:

$$\cdot \parallel \cdot \vdash \nabla[\texttt{exp}, \texttt{exp}](\lambda\{u, v\}[].\diamond, u : \texttt{P}, v : \texttt{pred}(\text{`}u) \gg \texttt{pred}(\text{`}u)) \in_{\mathbf{wf}} \texttt{jdg}$$

*Remark* 6.1. In the full development of a theory of behavioral refinements, the ad-hoc judgment $\Upsilon \parallel \Gamma \vdash M \in_{\mathbf{wf}} \tau$ would be replaced with a general notion of refinement $\phi \sqsubset \tau$ ($\phi$ refines sort $\tau$), and a new judgment $\Upsilon \parallel \Phi \vDash M \in \phi$, such that $\phi$ refines $\tau$ and the refinement context $\Phi$ refines some sort context $\Gamma$. Furthermore, in addition to restricting membership, refinements might also be designed to coarsen equivalence as in [5].

## Acknowledgements

The first author wishes to thank Robert Harper for numerous conversations about abstract binding trees and symbolic parameters, and for his feedback on a draft of this paper; and Andy Pitts for his help in understanding multi-sorted nominal sets and atomic sheaves.

# Appendices

## A $\mathbb{I}[\mathbb{S}]$, Nominal Sets, and the Schanuel Topos

Pitts defines a category **Nom** of nominal sets in [19] which function equivalently to the sheaves that we considered above, in the case of $\mathbb{S} \cong \mathbb{1}$. Fix a countably infinite set of atoms $\mathbb{A}$; then let $\mathbf{Perm}\,\mathbb{A}$ be the group of permutations (i.e. autoequivalences) on $\mathbb{A}$. Considered as a category, $\mathbf{Perm}\,\mathbb{A}$ has a single object $\bullet$ with morphisms $\pi : \bullet \longrightarrow \bullet$ for each $\pi \in |\mathbf{Perm}\,\mathbb{A}|$. Then, the category of nominal sets **Nom** is the subcategory of $\mathbf{Set}^{\mathbf{Perm}\,\mathbb{A}}$ containing just the presheaves which satisfy a *finite support* condition.

The category **Nom** is equivalent to the category of covariant sheaves on $\mathbb{I}$ under the atomic coverage, called the Schanuel Topos $\mathbb{S} \triangleq \mathbf{Sh}\,(\mathbb{I}^{op})$ [19, 9]; equivalently, the Schanuel topos is the subcategory of $\mathbf{Set}^{\mathbb{I}}$ containing only pullbacks-preserving functors.

In addition to the unisorted nominal sets framework, Pitts also briefly discusses a multi-sorted version of the apparatus in which the countably infinite set of atoms $\mathbb{A}$ is equipped with a sort assignment $\mathfrak{s} : \mathbb{A} \to \mathbb{S}$ such that all the fibers of $\mathfrak{s}$ are countably infinite; then, he defines for any such assignment the category $\mathbf{Nom}_{\mathfrak{s}}$ of $\mathfrak{s}$-sorted nominal sets as a subcategory of the category of presheaves on the group of $\mathfrak{s}$-respecting permutations on $\mathbb{A}$, $\mathbf{Perm}_{\mathfrak{s}}\,\mathbb{A}$.

In the same way as the Schanuel topos is equivalent to the unisorted nominal sets, we expect to find an equivalence between $\mathbf{Sh}\,(\mathbb{I}[\mathbb{S}])$ and the limit $\int_{\mathfrak{s}} \mathbf{Nom}_{\mathfrak{s}}$. At the very least, the sheaf topos $\mathbf{Sh}\,(\mathbb{I}[\mathbb{S}])$ may serve as a multi-sorted generalization of the Schanuel topos, for which we may carry over certain useful results, including its characterization as the subcategory of $\mathbf{Set}^{\mathbb{I}\downarrow\mathbb{S}}$ which contains just the pullbacks-preserving functors.

**Theorem A.1.** *A presheaf* $X : \mathbf{Set}^{\mathbb{I}\downarrow\mathbb{S}}$ *is a sheaf on* $\mathbb{I}[\mathbb{S}]$ *just when it preserves pullbacks.*

*Proof.* The proof is essentially the same as that of the analogous lemma for the Schanuel topos as presented in [14, A.2.1.11.h], but we will give a slightly more detailed version here. The presheaf X is a sheaf on the atomic site $\mathbb{I}[\mathbb{S}]$ just when, for any renaming $\rho : C \hookrightarrow D$ and any $n \in X(D)$ if $n \cdot \rho_0 = n \cdot \rho_1$ for all diagrams

$$C \overset{\rho}{\hookrightarrow} D \underset{\rho_1}{\overset{\rho_0}{\rightrightarrows}} E$$

such that $\rho_0 \circ \rho = \rho_1 \circ \rho$, then there exists a unique $m \in X(C)$ such that $m \cdot \rho = n$. In other words, the arrow $X(\rho)$ is an equalizer, as in the following:

$$\begin{array}{ccccc} X(C) & \xrightarrow{X(\rho)} & X(D) & \underset{X(\rho_1)}{\overset{X(\rho_0)}{\rightrightarrows}} & X(E) \\ {\scriptstyle m}\uparrow & \nearrow_{n} & & & \\ \mathbb{1} & & & & \end{array}$$

Now, because **Set** is a regular category, $X(\rho)$ is an equalizer precisely when it is a monomorphism. Therefore, if we have assumed that X preserves pullbacks, in order to demonstrate that X is a sheaf, it suffices to show that $X(\rho)$ is monic. If X preserves pullbacks, then it also preserves monomorphisms, because in any category with pullbacks, $f : A \longrightarrow B$ is monic just when the square

$$\begin{array}{ccc} A & \xrightarrow{1_A} & A \\ \downarrow{\scriptstyle 1_A} & & \downarrow{\scriptstyle f} \\ A & \xrightarrow{f} & B \end{array}$$

is a pullback [16, p. 16]. Because X preserves monomorphisms, and all arrows in $\mathbb{I} \downarrow \mathbb{S}$ are monic, then $X(\rho)$ is monic. Hence, X is a sheaf.

Next, we must show that if X is a sheaf, then it preserves pullbacks; we will loosely track the proof of Proposition 2.6.15 in [2]. Fix the following diagram such that it is a pullback square in $\mathbb{I} \downarrow \mathbb{S}$:

$$\begin{array}{ccc} C & \xrightarrow{f_j} & C_j \\ \downarrow{\scriptstyle f_i} & & \downarrow{\scriptstyle s_j} \\ C_i & \xrightarrow{s_i} & D \end{array}$$

Then it suffices to show that the following diagram is also pullback square in **Set**:

$$\begin{array}{ccc} X(C) & \xrightarrow{X(f_j)} & X(C_j) \\ \downarrow{\scriptstyle X(f_i)} & & \downarrow{\scriptstyle X(s_j)} \\ X(C_i) & \xrightarrow{X(s_i)} & X(D) \end{array}$$

We have to show that $X(C)$ is, up to equivalence, the pullback of $X(s_i)$ along $X(s_j)$ in **Set**, namely:

$$X(C_i) \times_{X(D)} X(C_j) \cong \{ (m_i, m_j) \in X(C_i) \times X(C_j) \mid m_i \cdot s_i = m_j \cdot s_j \}$$

The first direction of the equivalence is trivial: for each $m \in X(C)$, we can easily exhibit $(m \cdot f_i, m \cdot f_j) \in X(C_i) \times_{X(D)} X(C_j)$.

For the other direction, fix $(m_i, m_j) \in X(C_i) \times_{X(D)} X(C_j)$. Consider the singleton cover on C that contains $h \triangleq s_i \circ f_i = s_j \circ f_j$; a matching family $\phi$ for the cover $\{h\}$ is a map that takes h to an object in $X(D)$, and an amalgamation for $\phi$ consists in an object $m \in X(C)$ such that $\phi(h) = m \cdot h$. Now recall that X is a sheaf on $\mathbb{I}[\mathbb{S}]$ if and only if every matching family has a unique amalgamation; for the matching family $\phi(h) \triangleq m_i \cdot s_i = m_j \cdot s_j$, then, we must have a unique amalgamation $m \in X(C)$ since $X$ is a sheaf. Therefore, $X$ preserves pullbacks.

□